# A High-performance Atmospheric Radiation Package: with applications to the radiative energy budgets of giant planets

Cheng Li[1,*], Tianhao Le[1], Xi Zhang[2], Yuk L. Yung[1]

[1]California Institute of Technology

[2]University of California, Santa Cruz

* Corresponding author: cli@gps.caltech.edu

# Abstract

A High-performance Atmospheric Radiation Package (HARP) is developed for studying multiple-scattering planetary atmospheres. HARP is an open-source program written in C++ that utilizes high-level data structure and parallel-computing algorithms. It is generic in three aspects. First, the construction of the model atmospheric profile is generic. The program can either take in an atmospheric profile or construct an adiabatic thermal and compositional profile, taking into account the clouds and latent heat release due to condensation. Second, the calculation of opacity is generic, based on line-by-line molecular transitions and tabulated continuum data, along with a table of correlated-$k$ opacity provided as an option to speed up the calculation of energy fluxes. Third, the selection of the solver for the radiative transfer equation is generic. The solver is not hardwired in the program. Instead, based on the purpose, a variety of radiative transfer solvers can be chosen to couple with the atmosphere model and the opacity model.

We use the program to investigate the radiative heating and cooling rates of all four giant planets in the Solar System. Our Jupiter's result is consistent with previous publications. Saturn has nearly perfect balance between the heating rate and cooling rate. Uranus has the least radiative fluxes because of the lack of $CH_4$ and its photochemical products. Both Uranus and Neptune suffer from a severe energy deficit in their stratospheres. Possible ways to resolve this issue are discussed. Finally, we recalculate the radiative time constants of all four giant planet atmospheres and find that the traditional values from (Conrath BJ, Gierasch PJ, Leroy SS. Temperature and Circulation in the Stratosphere of the Outer Planets. Icar. 1990;83:255-81) are significantly overestimated.

1. **Introduction**

Many atmospheric radiative transfer programs have been developed for a specific purpose or for a specific planet. For example, the RFM (reference forward model) was developed to provide reference spectra for the infrared spectrometer on the Envisat satellite [1]; the SARTA (Stand-alone AIRS Radiative Transfer Algorithm) was designed to support the AIRS mission [2]; the NEMESIS (Non-linear Optimal Estimator for MultivariatE Spectral analySIS) planetary atmospheric radiative transfer model [3] was built to interpret the observations of Saturn and Titan from CIRS (Composite Infrared Spectrometer) onboard Cassini. Later on, their functionalities have been extended beyond their original purposes but still limited to the study of a specific class of planets. Adapting this type of radiative transfer model to an entirely different planet is possible but laborious and error-prone due to various hard-wired functions/assumptions intrinsic to the original design.

In the era of high precision spectroscopy on exoplanets enabled by future telescopes such as JWST, there is a pressing need to develop a generic radiative transfer program that is capable of computing atmospheric radiances and energy fluxes for exoplanets with arbitrary atmospheric composition and the presence of exotic clouds (e.g. [4]). Such a radiation model should be validated against the knowledge of Earth and other planets in the solar system where the observational constraints are available. Moreover, most radiation models used in planetary and exoplanetary studies are based on the models developed in the 1990s (e.g. [5, 6]). Although they are still working well, their software structures and efficiencies are behind the current industrial standard. Upon these considerations, it is the motivation behind the development of the new radiative transfer model.

Here we provide a new, open-source radiation model for Earth, planets and exoplanets that can calculate both line-by-line spectra and correlated-$k$ energy fluxes in parallel. The line-by-line spectra calculation is validated against the existing model (SARTA) developed for the Earth's satellite (AIRS) and the analytical solution published in Zhang, Nixon [7]. The correlated-$k$ opacities are generated using the line-by-line opacities. Depending on the backend radiative transfer equation solver, multiple scattering due to clouds and polarization can be accounted for. This paper presents the clear sky results for giant planets.

Validation of the line-by-line calculations against SARTA and cloud radiative effects will be discussed in a companion paper for the Earth.

The rest of this paper is organized as follows. In section 2, we discuss the key components of the model, including the atmospheric model, the opacity model, and the backend radiative transfer equation solver. In section 3, we apply the model to the giant planets in the solar system and calculate their heating/cooling rates and radiative time constants. The radiative forcing for ice giants are discusses in detail. Finally, in section 4, we summarize our major findings.

## 2. Overview of the implementation of HARP

HARP consists of three modular components: 1) An atmospheric model that either reads in or constructs a 1D atmosphere with thermal and compositional structure. The program is able to construct a (moist) adiabatic profile using the formulation of Li, Ingersoll [8], which allows the simultaneous condensation of multiple species and can handle any amount of condensable species, from zero to infinity. 2) An opacity model that calculates line-by-line optical properties (absorption coefficients, single scattering albedo, and coefficients in Legendre polynomial expansions of phase functions) for every layer of the atmosphere from a given atmospheric profile. The correlate-$k$ method is also provided as an option to further boost the calculation of radiative heating/cooling fluxes. 3) A backend radiative transfer model that takes the optical properties of the atmosphere computed in the opacity model as input and solves the radiative transfer equation. All three models are able to run in a stand-alone fashion, but they can also be combined to perform an efficient radiative transfer calculation from first principles. The program is written in a modern computer language (C++) that utilizes high-level data structure and algorithms. The code style conforms to the C++(11) standard, avoiding all language extensions. With efficiency being the priority, the opacity model and the radiative transfer model are executed in parallel for each atmospheric band, which significantly accelerates the speed of the program. On average, the program can compute one million eight-stream radiative transfer calculations in less than 10 mins on 15 Intel Xeon E5-2695 CPUs. The scaling using multiple cores is almost linear, meaning that using 15 CPUs will speed up the processing time by 15 times. Details of the numerical method used in each component are presented in the Appendix A.

**3. Heating rates and radiative time constants on giant planets**

The radiative heating/cooling rates of Jupiter's stratosphere has been calculated and discussed by many authors (e.g. [7, 9, 10]). Surprisingly, detailed analysis of the radiative energy budget has scarcely been done for Saturn, Uranus and Neptune. To complete the picture, we calculate the radiative heating/cooling rates and the relaxation timescales for all giant planets in the solar system in a single framework. The heating flux is averaged globally and annually, using the temperature profiles and hydrocarbon profiles from the state-of-the-art photochemical models. Those profiles are collected in Figure 1 and Figure 2 for comparison. Jupiter, Uranus and Neptune's profiles are from Moses, Fouchet [11]. Saturn's profiles are from Moses, Lellouch [12]. In all cases, 200 vertical computational layers are evenly spaced between 6000 and $10^{-2}$ hPa. Ring shadowing effects for Saturn's atmosphere has been considered according to the formulism and opacities given in Guerlet, Spiga [13]. The maximum ring shadowing effect occurs at 10º to 40º latitudes, reducing the solar insolation by about 25% [14].

The entire solar and infrared spectra are divided into 15 bands according to the absorption bands for $CH_4$, $C_2H_2$ and $C_2H_6$. The wavenumber range and the dominant absorber in each band is described in Kuroda, Medvedev [9], with spectral resolution of 0.01 $cm^{-1}$. We have considered the absorptions and emissions by $CH_4$, $C_2H_2$, $C_2H_6$, $NH_3$, $PH_3$ and the collision-induced absorptions by $H_2$-$H_2$ and $H_2$-He [15]. Molecular absorption lines are taken from HITRAN 2016 [16]. Cloud and aerosol opacities are not considered in the current study although they may have considerable contribution to the radiative forcing of the planet. Further discussions on the clouds and aerosol radiative effects are provided in Section 3.2.

It is worth noticing that the pressure broadening properties of hydrocarbons in a hydrogen-dominated atmosphere is different from those in a nitrogen-dominated atmosphere. The coefficients in the HITRAN database should be modified accordingly to account for the differences. We used the same modified coefficients as that in Zhang, Nixon [7], in which the $C_2H_6$ broadening coefficients are from Orton, Yanamandra-Fisher [17], the $C_2H_2$ broadening coefficients are from a parameterized fit to the results of Varanasi [18] and the $CH_4$ broadening coefficients are similar to those in the Earth's atmosphere [19].

In additional to the line-by-line calculations, we also generated correlated-*k* opacity tables for each planet and displayed them in Figure 3. For Jupiter, Uranus and Neptune, 16 Gaussian quadrature points are used for each interval within a spectral band. For Saturn, however, we found that 16 quadrature points are not sufficient to accurately capture the opacity distribution in the 960-2100 $cm^{-1}$ band. So, we have increased the number of quadrature points to 50 for Saturn. The correlated-k results are compared to line-by-line results in Figure 4. The radiative heating/cooling rates for Jupiter and Saturn are discussed in Section 3.1 while Uranus and Neptune are discussed in Section 3.2.

### 3.1. Jupiter and Saturn

Figure 4 (a) and (b) show the heating/cooling rates in the atmospheres of Jupiter and Saturn. Similar to the work by Zhang, Nixon [7] and others, Jupiter's heating rates increase with altitude and Saturn is alike. In general, the trend of the cooling rate follows that of the heating rate, however, near 200 hPa, the cooling rate in the infrared band (10-960 $cm^{-1}$) significantly decreases (blue line in Figure 4a) and even becomes positive (turns into heating) in Jupiter's atmosphere. This is caused by the strong temperature inversion occurred at the tropopause, where the warm air at both above and below heats the middle. Such effect also presents in Saturn's atmosphere but with a smaller magnitude and at a lower altitude. The correlated-*k* and the line-by-line methods give almost identical results. Their differences in the solar band (2100-9300 $cm^{-1}$) are generally less than 2% although large fractional differences should be expected when the heating/cooling rate approaches zero. In the infrared bands, the differences are slightly larger because the contributions of different molecules to the cooling rates vary with height, which impairs the correlation of absorption coefficients at different layers. Overall, the agreement between the line-by-line and the correlated-*k* models are within 10%.

The total heating/cooling rates are decomposed into contributions from individual gases in Figure 5. Jupiter and Saturn's stratospheric cooling is dominated by $C_2H_6$ above ~10 hPa level, and by the collision-induced absorptions (CIA) of $H_2$ below. Similar to the results of Yelle, Griffith [10] and Zhang, West [20], the cooling rates are slightly larger than the heating rates in Jupiter's stratosphere (Figure 5b), suggesting the existence of other heating sources, most likely the stratospheric haze particles [20]. Unlike Jupiter, Saturn's stratospheric heating rate almost balances its cooling rate (Figure 5d). If the missing of heating by haze particles is responsible for the energy deficit of Jupiter's stratosphere, the balanced energy budget of Saturn's stratosphere indicates that Saturn has less stratospheric haze compare to Jupiter. Roman, Banfield [21] analyzed high-resolution Cassini/ISS images with wavelengths from the ultraviolet to the near infrared. They concluded that the optical thickness of Saturn's stratospheric haze was very low ($\tau \sim 0.08$), a result that is consistent with our energy budget analysis.

### 3.2. Uranus and Neptune

Figure 4 (c) and (d) show the heating/cooling rates in the atmospheres of Uranus and Neptune. Uranus' heating/cooling rates are much smaller than those of other giant planets because Uranus has the coldest atmosphere (Figure 2) and has the least amount of stratospheric $CH_4$ gas among all giant planets (Figure 1). Because of the lack of $CH_4$ and the photochemical products thereof, the cooling of Uranus' atmosphere is almost solely due to the collision-induced absorption of $H_2$ (Figure 6a). By contrast, Neptune's heating/cooling rates are similar to those of the gas giants except that Neptune's stratosphere suffers from a major energy deficit at all pressure levels. The atmospheric cooling rate above the 0.2 hPa pressure level is almost two orders of magnitude larger than its heating rate. These findings are on par with the previously reported "stratospheric energy crises", which states that a much warmer stratosphere was observed than that calculated from a radiative model [5, 22, 23].

Although we have neglected the contributions from aerosols and clouds to the radiative energy budget, the clear sky calculation still sheds light on the problem and suggests possible ways to reconcile it. First, Marley and McKay [5] concluded that the observed stratospheric hazes could not explain the warm stratosphere, based on the assumption of spherical Mie scattering particles. It is possible that fractal aggregates (e.g. [20]) might provide sufficient heating for the atmosphere, as in Jupiter's atmosphere. Because the primary focus of this work is to introduce and document a new generic radiation model, a systematic study of clouds and aerosol effects on giant planets shall be deferred to later works. Second, the current calculation is based on the photochemical models of Moses, Fouchet [11] which did not include comet dust ablation. Were the new photochemical model [24] used, an even more severe problem would arise for the ice giant because of the elevated cooling from oxygen-bearing species coming from the meteoroid ablation and/or comet impacts. Hence, the dust ablation model should be re-examined in view of the constraints from the radiative energy budget.

### 3.3. Radiative time constant

The radiative time constant is an important timescale that is associated with radiative forcing of the atmosphere. By comparing it to the orbital timescale, one can infer whether a planet experiences strong seasonal effects [25]. We update the radiative time constant calculation for all giant planets by perturbing the temperature profile at every level by 5 K. The time constant is normalized by the orbital period and shown in Figure 7. In order to compare to Conrath, Gierasch [25]'s original figure (Figure 2, referred as CG afterwards), the normalized time constant is multiplied by $2\pi$.

Similar to what Zhang, Nixon [7] has found, the radiative time constant for Jupiter is about $T_{orb}/2\pi$ in the lower stratosphere and decreases upwards. Saturn and Neptune are similar to Jupiter, but Neptune's radiative time constant is the smallest among those three planets above 1 hPa because of the temperature increase in Neptune's upper stratosphere (Figure 2). Higher in the stratosphere, the mole fraction of the coolants, $C_2H_2$ and $C_2H_6$, increases with altitude, which helps decreasing the radiative time constants at high altitudes.

Uranus is an outlier in the family of giant planets for its low $CH_4$ abundance in the stratosphere and weak radiative forcing. Its radiative constant gradually increases with height above the tropopause. Overall, all giant planets show their radiative time constants to be about $T_{orb}/2\pi$ near the tropopause, a value that is much smaller than that in CG. This result is puzzling since the temperature profile used in CG was similar to ours and their hydrocarbon abundances were far larger than ours. However, we find three possibilities to explain why CG might overestimate the radiative time scale. First, the CIA coefficients used in CG were based on Dore, Nencini [26], and we have used an updated version of the CIA coefficients compiled by

Orton, Gustafasson [15]. Because the ordinate of Dore, Nencini [26]'s Figure 1 (absorption coefficients) was mislabeled, it is not possible to compare which one is larger. Moreover, CG represented the cooling from the $H_2$-$H_2$ CIA as $Q_H$ in their equation 24. But, unfortunately, they did not articulate how they calculated this term in their paper. Second, the infrared cooling formula used in CG was a crude approximation of the radiative transfer equation (assuming that radiation emitted directly to space from the level of interest) and only three spectral bands were used for the cooling due to hydrocarbons. Instead, we have solved a full radiative transfer equation over the entire cooling range (10-2100 $cm^{-1}$). Therefore, we argue that the difference between our result and CG's is likely due to their loose treatment of the $H_2$-$H_2$ CIA and the hydrocarbon absorptions, by using an older version of the CIA coefficients or by the oversimplification of the radiative transfer equation.

## 4. Conclusion

We developed a new integrated radiative transfer model, HARP, which calculates radiances and energy fluxes in parallel. It is designed to be flexible, efficient and capable of performing both line-by-line calculations and correlated-k calculations. The agreement between the line-by-line model and the correlated-*k* model are within 10%. We have detailed its major components, including the atmospheric model, the opacity model and the radiative transfer equation solver.

We applied the model to study the radiative forcing of the giant planets in the solar system. We find that Jupiter, Saturn and Neptune are alike while Uranus is an outliner in the family. Uranus is unique because of its low $CH_4$ abundance and the lacking of all the associated hydrocarbons in the stratosphere. Thus, the $CH_4$ condensation has a profound effect on the radiative properties of the atmosphere. Saturn's heating rate almost balances its cooling rate, suggesting the lack of large amount of haze particles in the stratosphere. Uranus and Neptune both exhibit a large deficit of energy, known as "stratospheric energy crises" in the literature on radiative equilibrium calculations, especially in the high stratosphere. Neptune has the largest imbalance between the heating and cooling among all giant planets with cooling rate uniformly larger than heating rate all over the stratosphere. Since the observed stratospheric hazes could not explain the warm stratosphere based on the assumption of spherical Mie scattering particles, further study of fractal aggregates should be pursued. We suggest a re-examination of the photochemical models in view of the constraints from the energy budget.

Finally, we updated the radiative time constants calculation. We found that the radiative time constants at the tropopause for all giant planets are about $T_{orb}/2\pi$, a value that is much smaller than those of Conrath, Gierasch [25]. We suspect two reasons for the discrepancy. One is that Conrath, Gierasch [25] used an older version of the $H_2$-$H_2$ CIA coefficients. Second is that Conrath, Gierasch [25] oversimplified the radiative transfer equation by neglecting the inter-level heat exchanges and assuming only three cooling bands for the hydrocarbons.

# Appendix: Implementation details of HARP

1. **Atmospheric module**

   The atmospheric module reads in a realistic atmospheric profile or constructs a 1D theoretical profile. The realistic atmospheric profile is usually the output from observations or from comprehensive numerical models. The 1D theoretical profile consists of two parts, separated by the tropopause, which is defined as where the temperature gradient starts to deviate from the adiabatic temperature gradient. Above, where the radiative process dominates, the temperature profile is determined by radiative equilibrium. Below, where the convective process dominates, the atmospheric profile is constructed from top to bottom on pressure coordinates assuming an adiabatic lapse rate. Condensation of clouds and the latent heat release are considered in the construction. The construction process and the formula of adiabatic temperature gradient including latent heats are referred to Li, Ingersoll [8].

2. **Opacity module**

   a. **Line-by-line opacities**

   The opacity module returns the absorption coefficients, the single scattering albedo and the coefficients in Legendre polynomial expansions of phase functions for a homogeneous layer in the atmosphere. If the phase function is described by a tabulated function of angles, a separate code is used to convert the tabulated values into the Legendre polynomial coefficients. In the following paragraphs, absorption coefficients, single scattering albedo and phase functions are generally referred to as optical properties. Because the optical properties can come from various sources, in order to consolidate different representations into a single framework, we generate a table of absorption coefficients restricted to a multidimensional bounded parameter space $\Omega$, in which the optical properties are linearly interpolated over a Cartesian mesh, which tessellates $\Omega$. For example, the HITRAN database [16] provides the intensity, the position, and the Voigt line shape parameters for each local transition lines. We use the Reference Forward Model (RFM) [1] to read line lists from HITRAN database and calculate the absorption coefficients at each wavenumber by summing the contributions from all nearby spectral lines. The multidimensional bounded parameter space $\Omega$ consists of pressure ($p$), temperature ($T$) and mixing ratios ($X$). The absorption coefficient ($Y$) at each Cartesian grid ($i, j, k$) is written as:

$$Y_{i,j,k} = Y(X_i, T_j, p_k), \tag{A1}$$

For each homogeneous layer of the atmosphere characterized by $(X, T, P)$, the absorption coefficient $Y$ is:

$$\begin{aligned} Y(X,T,P) &= a_i Y(X_i, T, p) + a_{i+1} Y(X_{i+1}, T, p) \\ &= a_i \left( b_j Y(X_i, T_j, p) + b_{j+1} Y(X_i, T_{j+1}, p) \right) \\ &\quad + a_{i+1} \left( b_j Y(X_{i+1}, T_j, p) + b_{j+1} Y(X_{i+1}, T_{j+1}, p) \right) \\ &= a_i b_j \left( c_k Y(X_i, T_j, p_k) + c_{k+1} Y(X_i, T_j, p_{k+1}) \right) + \\ &\quad a_i b_{j+1} \left( c_k Y(X_i, T_{j+1}, p_k) + c_{k+1} Y(X_i, T_{j+1}, p_{k+1}) \right) + \\ &\quad a_{i+1} b_j \left( c_k Y(X_{i+1}, T_j, p_k) + c_{k+1} Y(X_{i+1}, T_j, p_{k+1}) \right) + \\ &\quad a_{i+1} b_{j+1} \left( c_k Y(X_{i+1}, T_{j+1}, p_k) + c_{k+1} Y(X_{i+1}, T_{j+1}, p_{k+1}) \right), \end{aligned} \tag{A2}$$

where $a_i, b_j, c_k$ are the interpolation coefficients:

$$\begin{aligned} a_i &= \frac{X_{i+1} - X}{X_{i+1} - X_i}, \quad a_{i+1} = \frac{X - X_i}{X_{i+1} - X_i} \\ b_i &= \frac{T_{i+1} - T}{T_{i+1} - T_i}, \quad b_{i+1} = \frac{T - T_i}{T_{i+1} - T_i} \\ c_i &= \frac{p_{i+1} - p}{p_{i+1} - p_i}, \quad c_{i+1} = \frac{p - p_i}{p_{i+1} - p_i}. \end{aligned} \tag{A3}$$

If $X$, $T$, or $p$ is outside of the bounded parameter space, a constant extrapolation is performed. The interpolation process is generalized to arbitrary dimensions and to other optical properties in the program.

**b. Correlated-*k* opacities**

Although HARP is programed to run in parallel, which significantly speeds up the line-by-line calculations, in some cases, rapid but less accurate calculation of spectral integrated fluxes and heating rates are needed. We implemented the correlate-$k$ method as an option to further boost the speed of calculation. Because the theory and limitations of the correlated-$k$ method are discussed by previous authors (e.g. [27-29]), we outline only the method that is utilized in the current model. The comparison of the correlated-$k$ model and the line-by-line calculation is shown in Section 3, where the heating and cooling rates of four giant planets in the Solar System are calculated using both methods.

In the current correlated-$k$ implementation, the whole spectra are divided into several spectral bands, which typically has one dominant absorber. The spectrally integrated radiative flux of the band $b$ absorbed/emitted by a single homogeneous layer is:

$$F_b = \int_{\nu_1}^{\nu_2} F(\nu) \mathrm{d}\nu, \tag{A4}$$

where $F(\nu)$ is monochromatic flux at wavenumber $\nu$, and $\nu_1, \nu_2$ are wavenumber limits of band $b$. The integration over wavenumber can be changed into the integration over absorption coefficients:

$$F_b = (\nu_2 - \nu_1) \int_0^\infty f(k) F(k) \mathrm{d}k$$
$$= (\nu_2 - \nu_1) \int_0^1 F(g) \mathrm{d}g, \tag{A5}$$

where $f(k)$ is the probability density function of absorption coefficient $k$, and $g(k)$ is its cumulative probability function. So far, no approximation has been introduced in calculating $F_b$.

The next step is to approximate equation (A5) using a quadrature rule, i.e. a weighted sum of function values at specified points within the domain of integration. The discrete version of equation (A5) is:

$$F_b \approx (\nu_2 - \nu_1) \sum_{i=1}^{n} w_i F(g_i), \tag{A6}$$

where $w_i$ is the weight at quadrature point $g_i$, and $n$ is the number of quadrature points. The choice of the quadrature points is at will but usually Gaussian quadrature points yield better approximation to the integral given a fixed number of quadrature points [30]. Goody, West [28] and Lacis and Oinas [29] suggested that dividing the spectral band into evenly spaced intervals in ln($k$) results in good accuracy. By default, we choose three intervals, representing weak lines, intermediate lines and strong lines. Their dividing quantiles are at about 0.2 and 0.9. Choosing more intervals is possible but requires more computational time and does not necessarily reduce the error. The discretization process is repeated for all atmospheric levels using the same quadrature points. If the absorption coefficients at different atmospheric levels are correlated, each quadrature point $g_i$ corresponds to a single wavenumber $v_i$ for every level. Otherwise the $g_i$ values for different levels will not correspond to the same wavenumber, which is the major source of error in the correlated-$k$ method and cannot be reduced by increasing the quadrature points.

Because a strict correlation of absorption coefficients is usually not satisfied in a nonhomogeneous atmosphere, except for a few theoretical opacity models, e.g. Elsasser band model [31] or single-line models, we chose a layer $l$ in the atmosphere which approximately represents the overall distribution of absorption coefficients, and calculate the quadrature points $g_i$ according to the distribution of the absorption coefficients in this layer. A general rule-of-thumb to select the layer $l$ is that this layer should be deep enough to resolve the pressure-broadened molecular lines but the Voigt line profiles in this band do not overlap too much. It is possible to run a series of models to find the best value of $l$ that gives the closest result to the line-by-line model. But after several trial-and-error attempts, we find that the rule-of-thumb provides sufficient accuracy.

A subtler and more esoteric aspect of the correlated-$k$ method is the treatment of the Planck function. Goody, West [28] assumed an average Planck function over the entire spectral interval, despite that Planck function is a strong function of temperature. Such simple approximation was used in various published correlated-$k$ models (eg. [13], [27]) and has achieved good accuracy if the spectral bands are carefully chosen. An improvement over the constant Planck function is the Planck-weighting scheme in which the absorption coefficients are weighted by the value of Planck function [29]. This method results in a very large interpolation table of absorption coefficients, weighted by Planck function and transmittance, in pressure, temperature and absorber abundance. Similarly, Amundsen, Tremblin [27] found that there is no significant difference between the weighted and unweighted method if the spectral bands are chosen appropriately for hot Jupiters' atmospheres. To balance the accuracy and the flexibility of the model, we find that a simple modification of the constant Planck function approximation could yield better accuracy at very small computational cost. The idea is to average the Planck function over the spectral lines with similar absorption coefficients. The modification looks alike the Planck-weighting scheme but lessen the burden of storing a large look-up table of temperature.

Suppose the quadrature points are $\{g_i \mid i = 1 \ldots n\}$ in equation (A6), satisfying $0 < g_i < 1, i = 1 \ldots n$. We extend the set $\{g_i\}$ by two boundary values: $g_0 = 0$ and $g_{n+1} = 1$. The average Planck function used at quadrature point $g_i$ is defined as:

$$\bar{B}(T, g_i) = \frac{1}{g_{i+1} - g_{i-1}} \int_{g_{i-1}}^{g_{i+1}} B(T, g) \mathrm{d}g, \qquad i = 1 \ldots n, \tag{A7}$$

where $B(T, g)$ is the Planck function at temperature $T$ and at the frequency $\nu$ that corresponds to the cumulative probability of $g$. Note that the frequency space has been changed to the probability space. In the continuous limit, where $n \to \infty$, $\bar{B}(T, g_i) = B(T, g_i)$. The average Planck function is identical to the individual one. Conversely, if there is only one quadrature point, $\bar{B}(T, g_i) = \int_0^1 B(T, g) \mathrm{d}g$ is the average Planck function over the entire spectral interval, which is the same as the constant Planck function approximation. This scheme is similar to the usual "three-point moving average" method.

Figure A1 illustrates why the "moving averaging" scheme may improve the accuracy of the correlated-$k$ method. The left panel in Figure A1 shows absorption coefficients of $H_2O$ and the incoming solar radiation in the near-infrared band (5150-6150 cm$^{-1}$). The incoming solar radiation in general increases with wavenumber, but the absorption coefficients exhibit a decreasing trend. It implies that solar flux is smaller for stronger lines and larger for weaker lines. If a constant Planck function were used, the heating by strong lines would be overestimated and heating by weak lines would be underestimated. Using the "moving averaging" scheme proposed by equation (A7) correctly captures the overall trend of the variation of the Planck function in the spectral range (right panel in Figure A1). Nevertheless, a quantitative evaluation of how much accuracy does the averaging gain compared to a constant Planck function or a Planck-weighted method is hard to establish because the error introduced by the variation of Planck function is usually subordinated to the error introduced by the assumption of spectral correlation. Choosing the spectral bins in a smart way, as stated in [13] and [27], or constructing an immense look-up table of absorption coefficients, as the case of RRTM, can both improve the accuracy of the correlated-$k$ method. As a result, which weighting method to use is determined by the accuracy requirement. We favor such a "moving averaging" scheme because it is simple to implement, results in good accuracy and is relatively insensitive to the choice of spectral bins.

c. **Radiative transfer module**

A variety of computational algorithms for solving the radiative transfer equations are available. We demonstrate using the C-version of DISORT program [32] as the solver for the purpose of calculating the radiance and flux of a layered media. Using the C-version of the code has several advantages over the original Fortran version. First, because our program is written in C++, interfacing to a radiative transfer solver written in C is seamless. Second, the C-version uses dynamic and cache-aware memory allocation, which speeds up the program significantly without sacrificing accuracy [32]. For a Fortran user, interfacing to a Fortran radiative transfer program should not take too much effort because C++ is a compiled higher-level language than Fortran. The detailed steps are provided in various standard materials, such as *Fortran 77 Programmer's Guide*.

The optical thickness of each atmospheric layer is obtained by multiplying the absorption coefficient by the path length $\Delta s$:

$$\Delta \tau = Y(\bar{X}, \bar{T}, \bar{P}) \Delta s$$
$$= Y(\bar{X}, \bar{T}, \bar{P}) \frac{\Delta p}{\rho(\bar{X}, \bar{T}, \bar{P}) g} \sec \theta , \quad (A8)$$

where the $\overline{(\ )}$ symbol indicates the layer averaged quantity, $\rho(\bar{X}, \bar{T}, \bar{P})$ is the average density calculated from the equation of state, $g$ is the gravity, $\Delta p$ is the pressure difference between the top and the bottom of this layer, and $\theta$ is the viewing angle with respect to the vertical.

Given the optical thickness $\Delta \tau$ (defined at layers), single scattering albedo $\omega(\tau)$ (defined at levels), and the scattering phase function $P(\tau, \mu, \phi)$, the DISORT algorithm divides the monochromatic radiation field into 2N discrete streams and solves the discretized radiative transfer equation. For an absorbing atmosphere, we find almost no difference in radiative fluxes obtained from using 8 and 16 streams. For scattering atmospheres, usually 32-stream should be used. Note that the computational time scales with $N$ as $N^2$.

Given the modular structure of HARP, we have the option of using other radiative transfer solvers such as LIDORT [33] and 2S-ESS [34], allowing great flexibility. For example, LIDORT provides analytic Jacobians (needed in inversion studies) at an additional cost that is a fraction of the original computational cost. 2S-ESS separates the single-scattering component from multiple scattering.

## Acknowledgements

We thank Julie Moses for kindly providing the photochemical model results of four giant planets. X.Z. acknowledges support from NASA Solar System Workings grant NNX16AG08G. C.L. acknowledges the support from NASA Postdoc Program Fellowship.

## Code availability

HARP is an open-source program and will be made freely available to the community on Github at https://github.com/luminoctum/athena-harp.

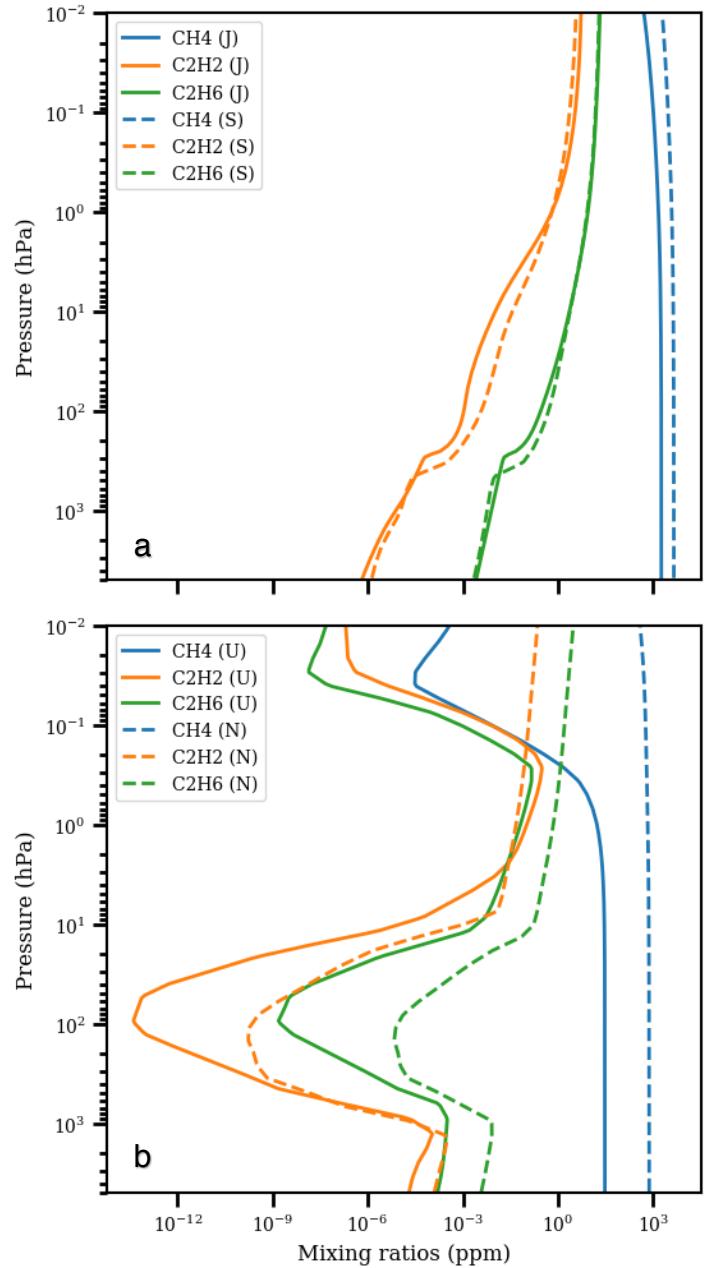

Figure 1. Vertical profiles of hydrocarbons from 1-D photochemical models. a) Jupiter and Saturn's profiles are from Moses, Fouchet [35] and Moses, Bezard [36]. b) Uranus and Neptune's profiles are from Moses, Fouchet [35].

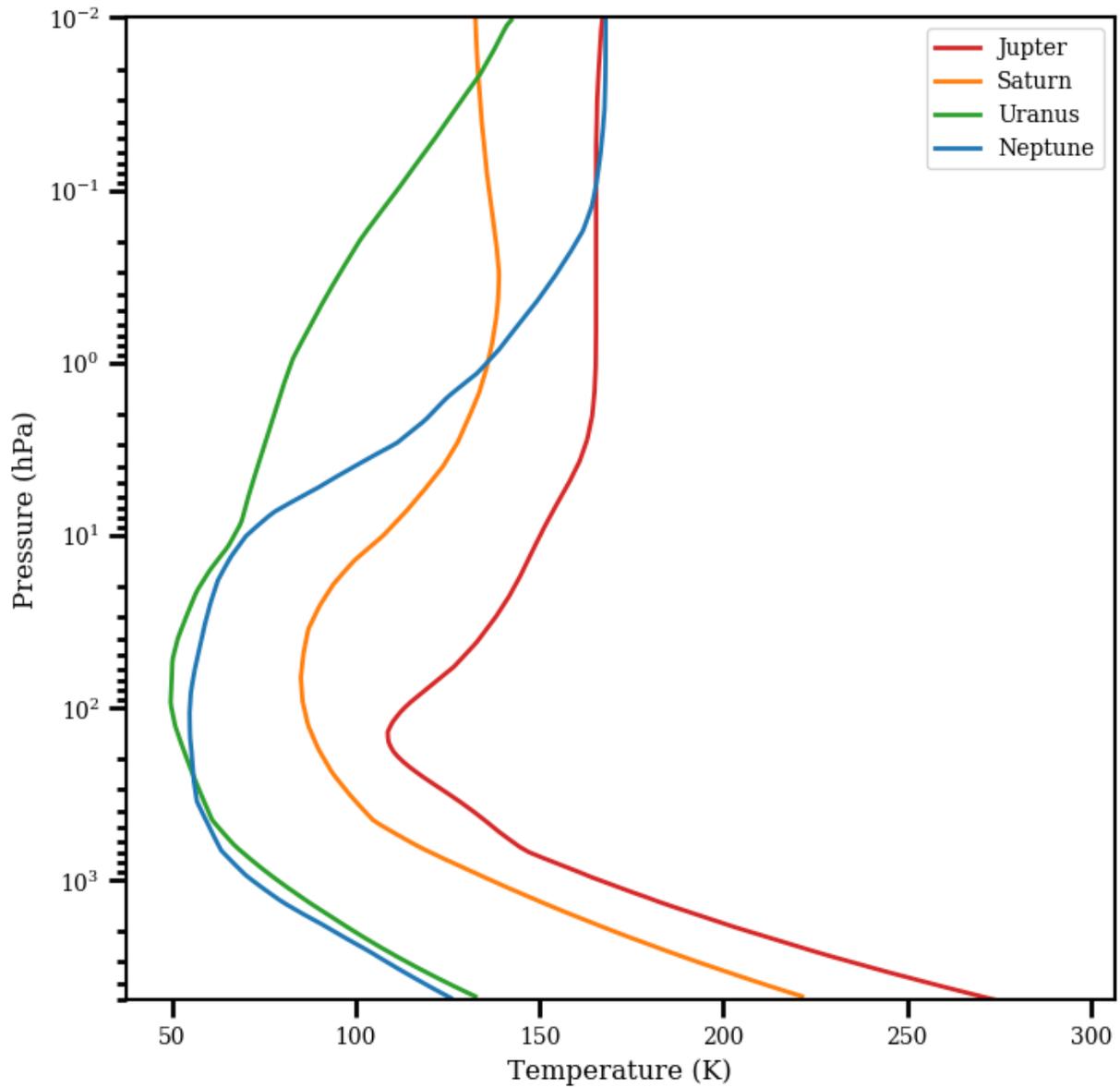

Figure 2. Temperature profiles from the stratosphere to the upper troposphere used in the model. Similar to the hydrocarbon profiles, these temperature profiles are from Moses, Bezard [36] and Moses, Fouchet [11].

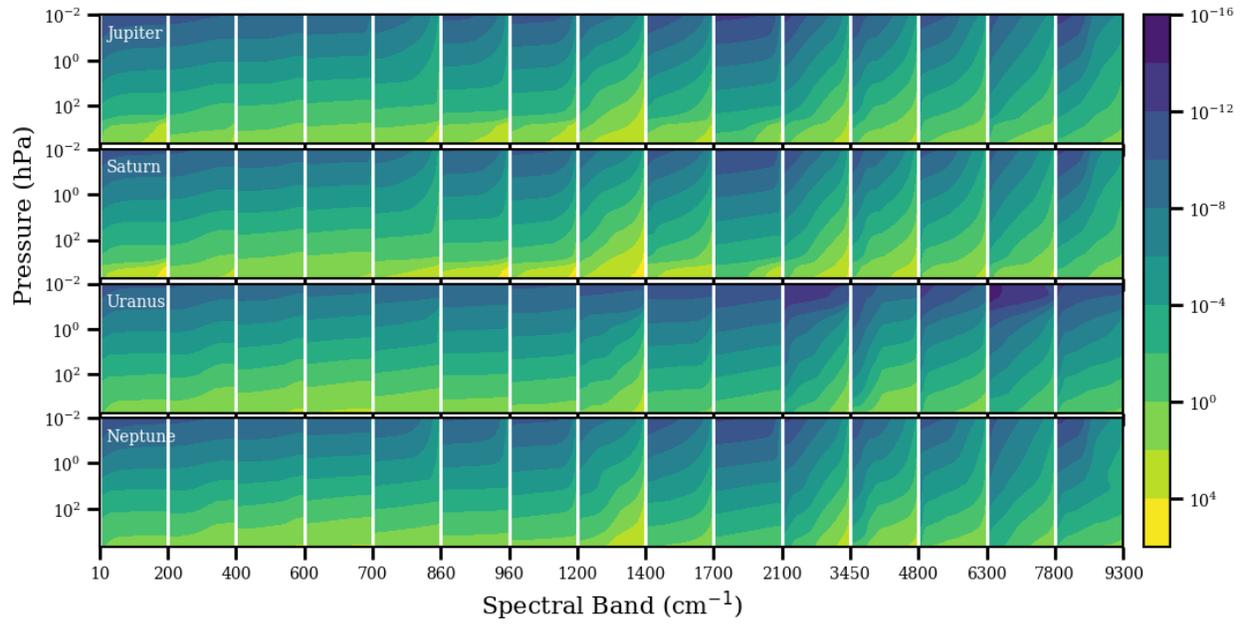

Figure 3. Correlated-*k* optical depths. Jupiter, Saturn, Uranus and Neptune are displayed from top to bottom. Boundaries of spectral bands are indicated in the abscissa. Within each spectral band, the absorption coefficients are sorted layer by layer in ascending order. The ordinate of each small box bounded by two vertical white lines indicates the pressure level and the abscissa of that represents the cumulative probability of absorption coefficients.

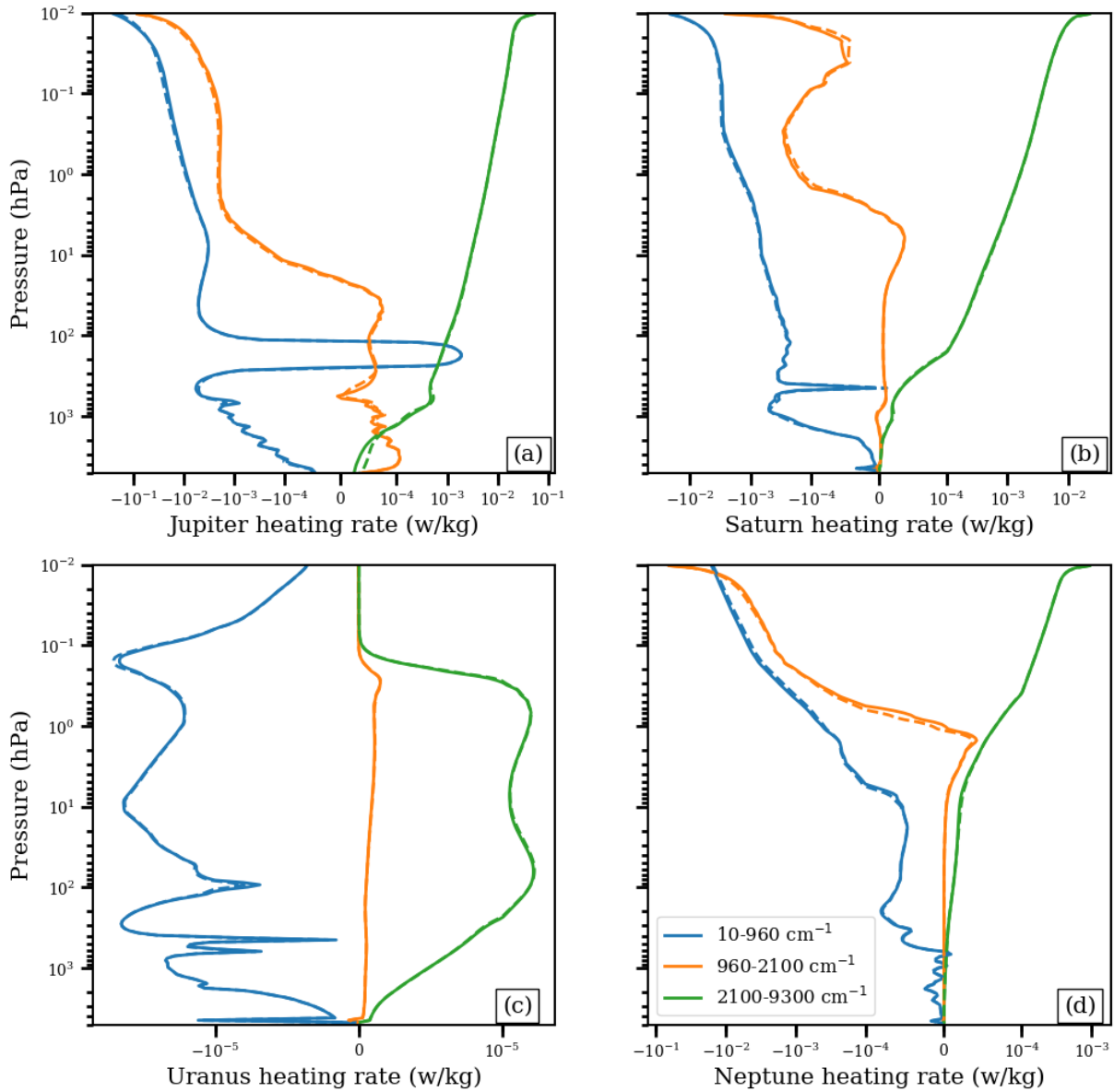

Figure 4. Heating rates in three spectral bands: 10-960 cm$^{-1}$ represents the thermal band (cooling); 960-2100 cm$^{-1}$ represents the thermal/solar mixed band (cooling & heating); 2100-9300 cm$^{-1}$ represents the solar band (heating). (a) Jupiter, (b) Saturn, (c) Uranus and (d) Neptune are ordered from left to right and top to bottom. Correlated-k results (solid lines) are compared with line-by-line calculations (dashed lines). Negative value of heating rate is cooling rate.

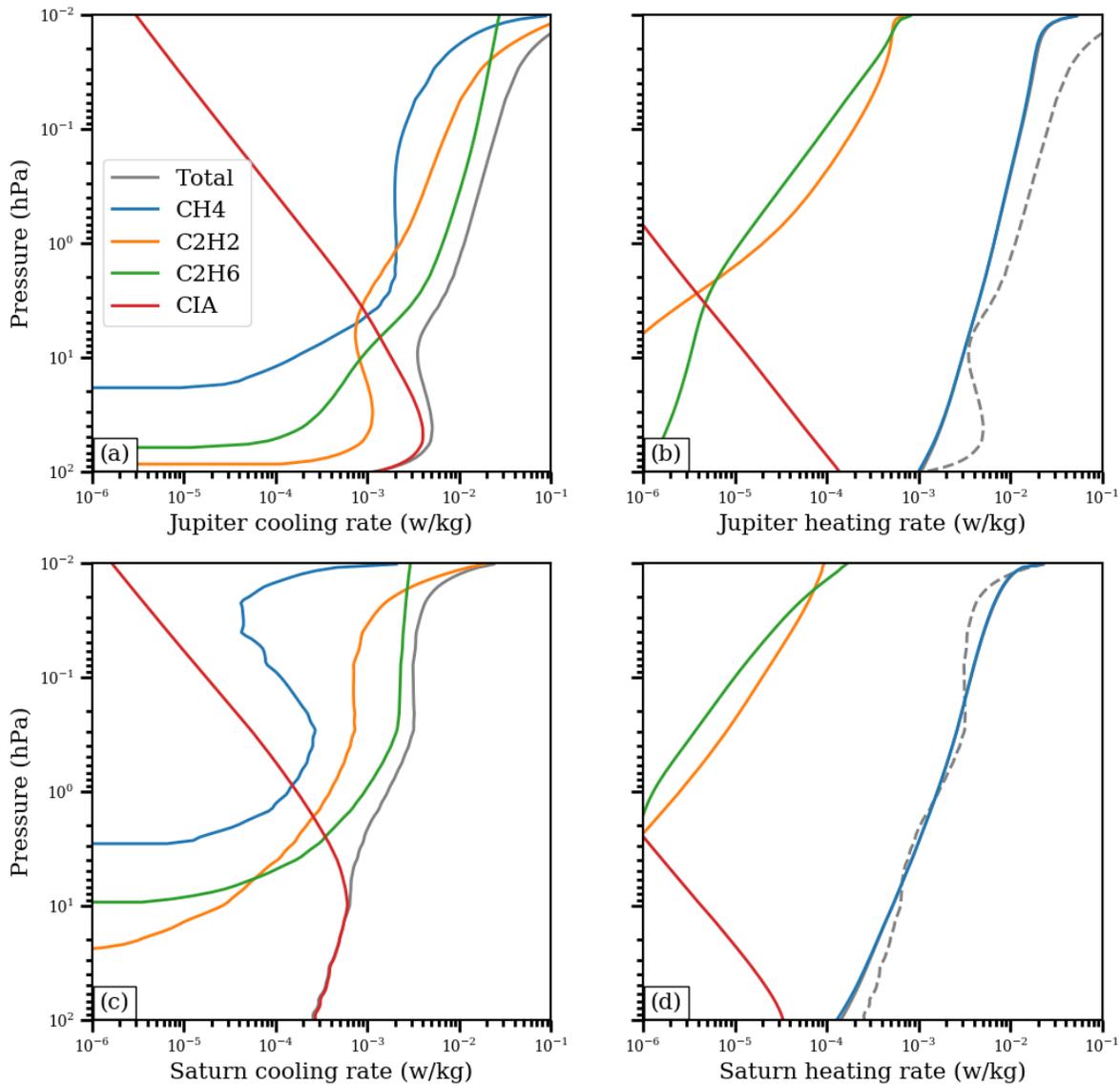

Figure 5. Contribution of different gases to the heating/cooling rates of Jupiter and Saturn. Heating rate is calculated by integrating the shortwave fluxes from 2100 to 9300 cm$^{-1}$. Cooling rate is calculated by integrating the longwave fluxes from 10 to 2100 cm$^{-1}$. The dashed lines in panels (b) and (d) mirrors the total cooling rates (the black line) in panels (a) and (c) to show the radiative energy budget. Note that the atmospheric heating is dominated by CH$_4$ absorption, and therefore the blue line (heating rate due to CH$_4$ absorption) is hidden beneath the black line (total heating rate) in panels (b) and (d).

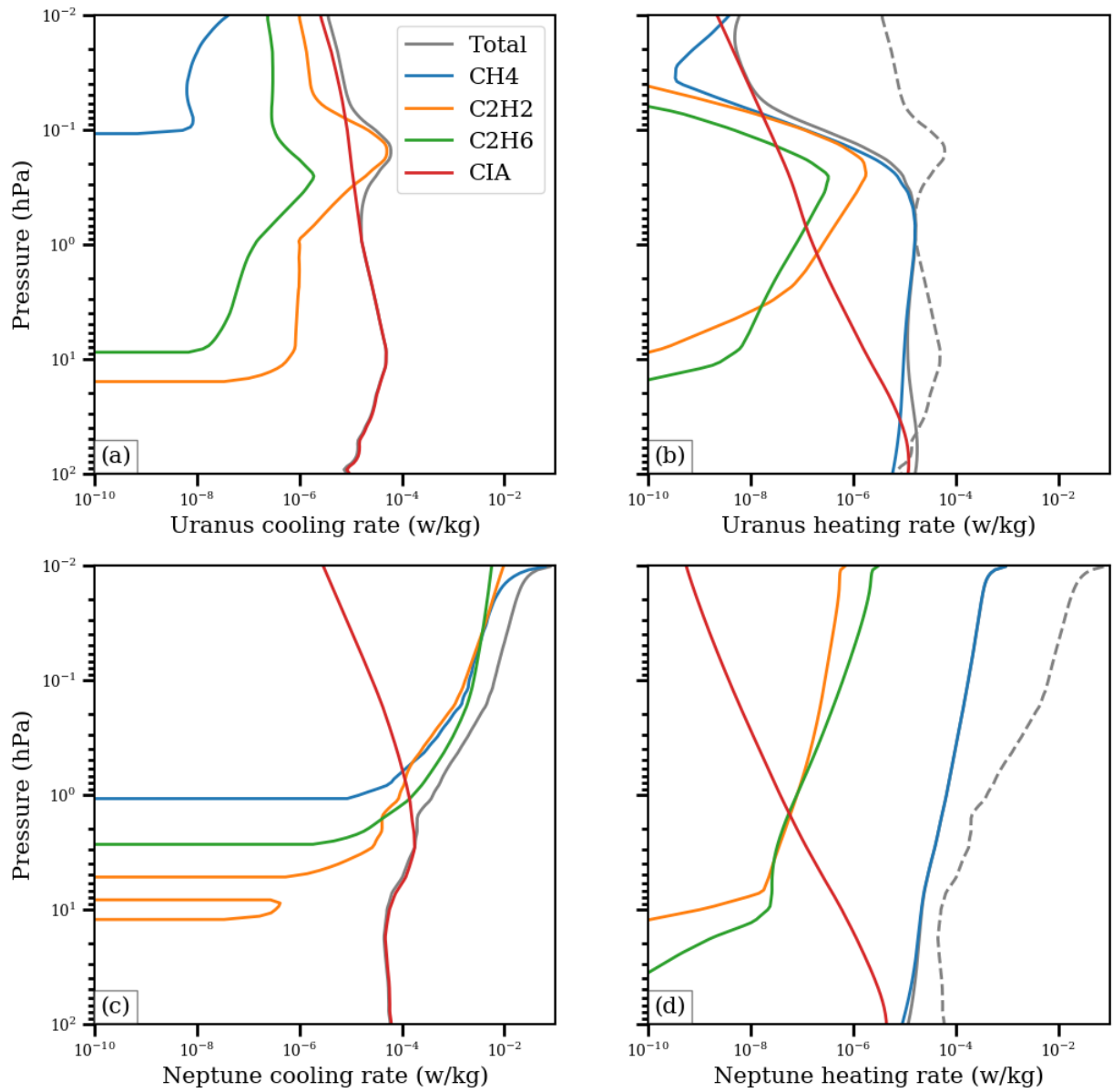

Figure 6. Same as Figure 5 but for Uranus and Neptune.

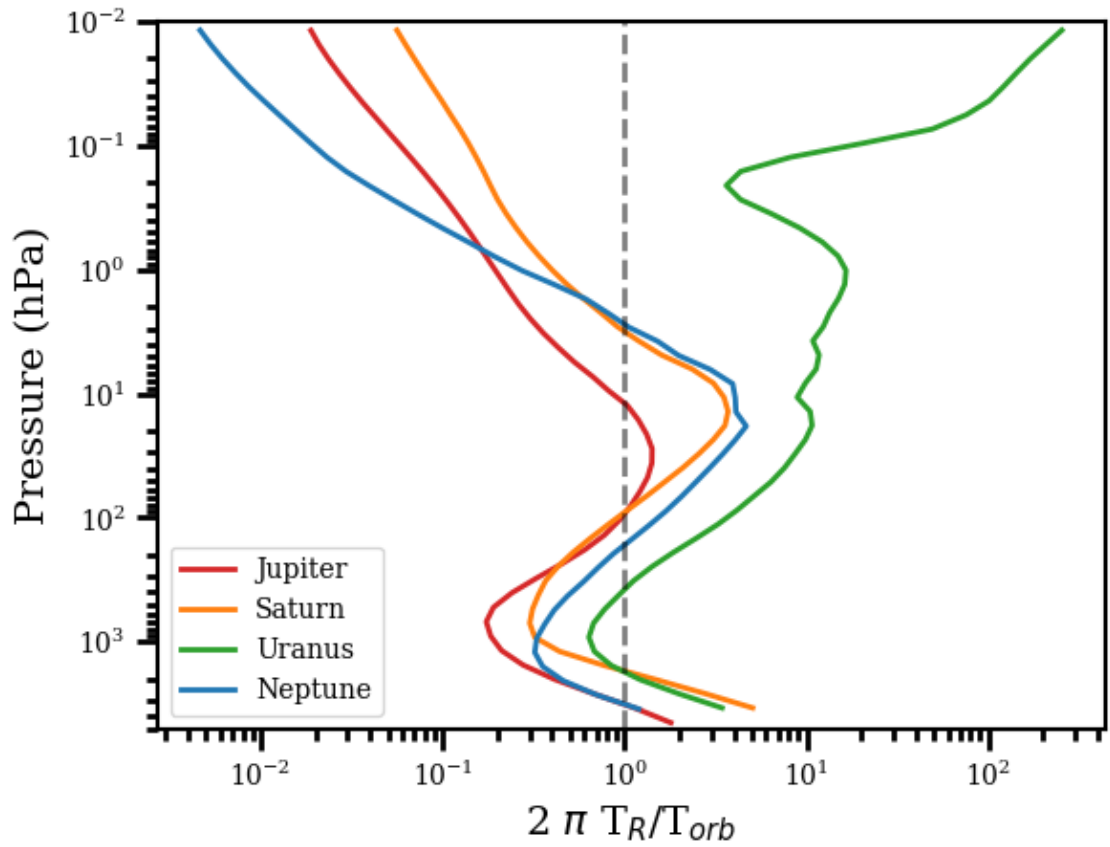

Figure 7. Vertical profiles of radiative time constant based on the temperature profiles in Fig. 2 and the cooling rate profiles in Fig. 4. The radiative time constant $T_R$ is normalized by the orbital period $T_{orb}$, which is 11.9, 29.5, 84 and 165 Earth years for Jupiter, Saturn, Uranus and Neptune, respectively.

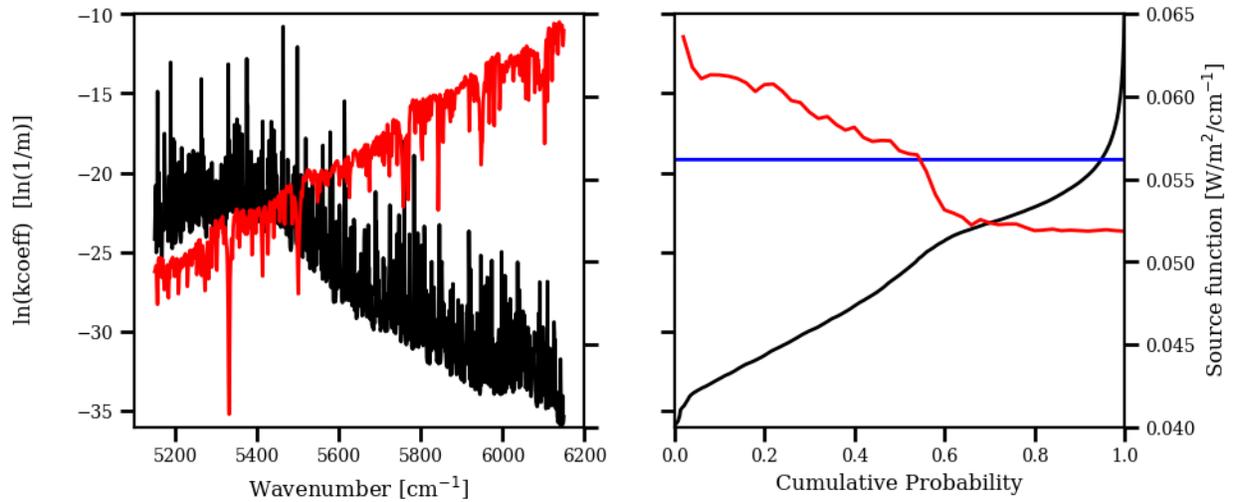

Figure A1. Left panel: solar radiation (red, y-axis on the right) and H$_2$O absorption coefficients (black, y-axis on the left) in a NIR spectral band (5150-6150 cm$^{-1}$). Right panel: Cumulative probability distribution $g(k)$ of the absorption coefficients ($k$) in the NIR band in the left panel after spectral regrouping (black). The averaged solar radiation according to equation (13) is shown in red. The average Planck function of the entire band is shown in blue.